\documentclass{llncs}
\usepackage{makeidx}  
\usepackage{graphicx} 
\usepackage{subeqnar} 
\usepackage{multicol} 

\usepackage{amsmath}
\usepackage{amssymb}
\usepackage{amscd}
\usepackage{latexsym}
\usepackage{epsfig}

\newtheorem{defi}{Definition}
\newtheorem{Definition}[defi]{Definition}
\newtheorem{Theorem}[defi]{Theorem}

\newtheorem{Lemma}[defi]{Lemma}

\date{}

\begin{document}

\pagestyle{plain}  

\title{A polynomial quantum query lower bound for the set equality problem}
\titlerunning{A polynomial quantum query lower bound for the set equality problem}

\author{Gatis Midrij\= anis
\thanks{Research supported by Grant No.01.0354 from the
Latvian Council of Science, and  Contract IST-1999-11234 (QAIP)
from the European Commission.}}
\authorrunning{Gatis Midrij\= anis }

\institute{University of Latvia, Rai\c na bulv\= aris 19, Riga,
Latvia. Email: \texttt{gatis@zzdats.lv}. Fax:
\texttt{+371-7820153}.}

\maketitle


\begin{abstract}
    The set equality problem is to tell whether two sets
    $A$ and $B$ are equal or disjoint under the promise that one
    of these is the case. This problem is related to the Graph Isomorphism
    problem. It was an open problem to find any $\omega(1)$
    query lower bound when sets $A$ and $B$ are given by quantum
    oracles. We will show that any error-bounded quantum query algorithm that solves
    the set equality problem must evaluate oracles $\Omega(\sqrt[5]{\frac{n}{\ln n}})$ times, where
    $n=|A|=|B|$.
\end{abstract}

\section{Introduction, motivation and results}
\label{sec:Intro}

The amazing integer factoring algorithm of Shor~\cite{ShorFact}
and search algorithm of Grover~\cite{Grover} show that to find
quantum lower bounds is more that just a formality. The most
popular model of quantum algorithms is the query (oracle) model.
Thus, also quantum lower bounds are proved in the query model.
There are developed methods that offer tight or nearly tight lower
bound for some problems, however for some other problems not.
Recently Aaronson~\cite{Aaronson} found a new method how to get
tight quantum query lower bounds for some important problems, for
example, the collision problem. This was an open problem since
1997. Aaronson's method uses symmetrization over the input and
therefore can be hard to apply to the problems with asymmetric
input. The set equality problem is an example of such problem and
it remaind unsolved.

In this paper we will find a quantum lower bound for the set
equality problem by reduction. We will reduce the collision
problem to the set equality problem, therefore getting quantum
query lower bound for the set equality problem.

Let assure ourselves that the set equality problem is related with
Graph Isomorphism problem. We are given two graphs $G_1, G_2$ and
we want to establish whether there exists permutations $p_1, p_2$
over vertices of graphs such that permutated graphs $p_1(G_1),
p_2(G_2)$ are equivalent (graphs $G_1, G_2$ are isomorphs). Let
$P_i$ denote the set of all graphs gotten by some permutation over
graph $G_i$'s vertices ($i \in \{0,1\}$). It is easy to see that
if graphs are isomorphs then $P_1 = P_2$, but if not, then $P_1
\cap P_2 = \O$. Therefore, if one can distinguish between those
cases, then he can solve the Graph Isomorphism problem. Since
there are $n!$ permutations for a graph with $n$ vertices , the
sizes of $P_1, P_2$ can be superpolynomial over the number of
vertices of graphs $G_1, G_2$.

Let $[n]$ denote the set $\{1, 2, ..., n\}$.

\begin{Definition}
Let $a:[n]\mapsto [N]$ and $b:[n]\mapsto [N]$ be the functions.
Let $A$ be the set of all $a's$ images $A := a([n]) := \{a(1),
a(2), ..., a(n)\}$ and $B := b([n]) := \{b(1), b(2), ..., b(n)\}$.
There is the promise that either $A = B$ or $A \cap B = \O$.

Let \textbf{the general set equality} problem denote the problem
to distinguish these two cases, if functions $a$ and $b$ are given
by quantum oracles.
\end{Definition}
By use of Ambainis'~\cite{Ambainis} method it is simple
(\cite{MansSetEq}) to prove $\Omega(\sqrt{n})$ lower bound for the
general set equality problem. However, this approach works only if
every image can have very many preimages. Graph theorists think
that the Graph Isomorphism problem, when graphs are promised not
to be equal with themselves by any nonidentical permutation, still
is very complex task. This limitation lead us to the set equality
problem where $a$ and $b$ are one-to-one functions.

\begin{Definition}
Let \textbf{one-to-one set equality} problem denote the general
set equality problem under promise that $a(i) \neq a(j)$ and $b(i)
\neq b(j)$ for all $i \neq j$.
\end{Definition}

Finding $\omega(1)$ quantum query lower bound for the set equality
problem was posed an open problem by Shi\cite{ShiColl}. Despite
$\omega(1)$ lower bound for the one-to-one set equality problem
remaining unsolved task, $\Omega(\frac{n^{1/3}}{\log^{1/3}n})$
quantum query lower bound was showed \cite{MansSetEq} for a
problem between these two problems when $|a^{-1}(x)| = O(\log n)$
and $|b^{-1}(x)| = O(\log n)$ for all images $x \in [N]$.

In this paper we will show the polynomial quantum query lower
bound for the most challenging task: the one-to-one set equality
problem.
\begin{Theorem}\label{thm:Main}
    Any error-bounded quantum query algorithm $\mathcal{A}$ that solves
    the one-to-one set equality problem  must evaluate functions $\Omega(\sqrt[5]{\frac{n}{\ln n}})$ times.
\end{Theorem}

The rest of the paper will be organized as follows. In the
section~\ref{sec:Preliminaries} will be notations and previous
results that we will use. In the section~\ref{sec:Idea} we will
preview the main idea of the proof of Theorem~\ref{thm:Main}. The
section~\ref{sec:Framework} will start the proof, the
section~\ref{sec:DistinctionPrep} will prepare for continuing
proof and the section~\ref{sec:Compl} will finish the proof.

\section{Preliminaries}
\label{sec:Preliminaries}

\subsection{Quantum query algorithms}
\label{sec:Preliminaries:upper}

The most popular model of the quantum computing is a query (or
oracle, or black box) model where the input is given by the
oracle. For more details, see a survey by Ambainis~\cite{AmbSurv}
or a textbook by Gruska~\cite{Gruska}. In this paper we are able
to skip them because our proof will be based on reduction to
solved problems.

In this paper we consider only the worst case complexity for
error-bounded quantum algorithms. Thus, without loss of
generality, we can assume that any quantum algorithm makes the
same number of queries for any input. If we say that algorithm has
two input functions $a$ and $b$ then for technical reasons
somewhere it can be comprehend with one input function denoted as
$(a,b)$.

One of the most amazing quantum algorithms is Grover's search
algorithm (\cite{Grover}). It shows how a given $x_1 \in \{0, 1\},
x_2 \in \{0, 1\}, ..., x_n \in \{0, 1\}$ to find the $i$ such that
$x_i = 1$ with $O(\sqrt{n})$ queries under promise that there
exists at most one such $i$.

This algorithm can be considerably generalized to so called
amplitude amplification \cite{AmplAmpl}. Using amplitude
amplification one can make good quantum algorithms for many
problems till the quadratic speed-up over classical algorithms.

By straightforward use of amplitude amplification we get quantum
algorithm for the general set equality problem making
$O(\sqrt{n})$ queries and quantum algorithm for the one-to-one set
equality problem making $O(n^{1/3})$ queries. Therefore our lower
bound probably is not tight.

\subsection{Quantum query lower bounds}
\label{sec:Preliminaries:lower}

There are two main approaches to get good quantum query lower
bounds. The first is Ambainis' \cite{Ambainis} quantum adversary
method, other is lower bound by polynomials introduced by Beals et
al. \cite{Beals} and substantially generalized by
Aaronson~\cite{Aaronson}, Shi~\cite{ShiColl} and others. Although
explicitly we will use only Ambainis' method, the lower bound we
will get by the reduction to the problem, solved by polynomials'
method.

The basic idea of the adversary method is, if we can construct a
relation $R \subseteq X \times X$, where $X$ and $Y$ consist of
0-instances and 1-instances and there is a lot of ways how to get
from an instance in $X$ to an instance in $Y$ that is in the
relation and back by flipping various variables, then query
complexity must be high.

\begin{Theorem}
\label{AThm} \cite{Ambainis} Let $f(x_1, ..., x_n)$, be a function
of n variables with values from some finite set and $X, Y$ be two
sets of inputs such that $f(x) \neq f(y)$ if $x \in X$ and $y \in
Y$. Let $R \subset X \times Y$ be such that
\begin{itemize}
    \item For every $x \in X$, there exist at least $m$ different $y
        \in Y$ such that $(x, y) \in R$.
    \item For every $y \in Y$, there exist at least $m'$ different
    $x \in X$ such that $(x, y) \in R$.
    \item For every $x \in X$ and $i \in \{1, ..., n\}$, there are at most $l$ different
    $y \in Y$ such that $(x, y) \in R$ and $x_i \neq y_i$.
    \item For every $y \in Y$ and $i \in \{1, ..., n\}$, there are at most $l'$ different
    $x \in X$ such that $(x, y) \in R$ and $x_i \neq y_i$.
\end{itemize}
Then, any quantum algorithm computing $f$ uses
$\Omega(\sqrt{\frac{m m'}{l  l'}})$ queries.
\end{Theorem}
Actually, original Ambainis' formulation was about $\{0,
1\}$-valued variables but we can use any finite set as it is
implied by the next, more general theorem in Ambainis'
paper~\cite{Ambainis}.

\subsection{The collision problem}
\label{sec:Preliminaries:Collision}

Finding $\omega(1)$ quantum lower bound for the collision problem
was an open problem since 1997. In 2001 Scott Aaronson
\cite{Aaronson} solved it by showing polynomial lower bound. Later
his result was improved by Yaoyun Shi \cite{ShiColl}. Recently,
Shi's result was extended by Samuel Kutin \cite{Kutin} and by
Andris Ambainis \cite{AmbCol} in another directions.

Below is an exact formulation of the collision problem due to
Shi\cite{ShiColl}.
\begin{Definition}
\label{def:CollShi} Let $n > 0$ and $r \geq 2$ be integers with
$r|n$, and let a function $f$ of domain size n be given as an
oracle with the promise that it is either one-to-one or r-to-one.
Let \textbf{r-to-one} collision problem denote the problem to
distinguishing these two cases.
\end{Definition}

Shi~\cite{ShiColl} showed following quantum lower bound for the
r-to-one collision problem.
\begin{Theorem} ~\cite{ShiColl} \label{shi}
    Any error-bounded quantum algorithm that solves r-to-one collision problem must
    evaluate the function $\Omega((n/r)^{1/3})$ times.
\end{Theorem}
Kutin \cite{Kutin} and Ambainis \cite{AmbCol} extended his result
for functions with any range.

\subsection{Notations}
\label{sec:Preliminaries:Notations}

Let $F^*:=F^*(n,N)$ denote the set of all partial functions from
$[n]$ to $[N]$. Then any $f^* \in F^*$ can be conveniently
represented as a subset of $[n] \times [N]$, i.e.,
$f^*=\{(i,f^*(i)):i \in dom(f^*)\}$.

For a finite set $K \subseteq \mathbf{Z}^+$, let $SG(K)$ denote
the group of permutations on $K$. For any integer $k>0$, $SG(k)$
is a shorthand for $SG([k])$. For each $\sigma \in SG(n)$ and
$\tau \in SG(N)$ define $\Gamma_\tau^\sigma:F^*\rightarrow F^*$ as
$$\Gamma_\tau^\sigma(f^*):=\{(\sigma(i),\tau(j)):(i,j)\in f^*\},
\forall f^* \in F^*.$$

\section{The idea behind the proof}
\label{sec:Idea} The rest of the paper is proof of the
Theorem~\ref{thm:Main}. In this section we will discuss the main
idea behind this proof. The key is to reduce some problem with
known quantum query lower bound to the one-to-one set equality
problem. Unfortunately, a simple reduction does not work.
Therefore we must make a chain of reductions and in the end get
the problem, which can be solved by arbitrary methods.

The problem, which we will try to reduce to the one-to-one set
equality problem, is the collision problem. All steps of reduction
will be probabilistic. One of that steps
Midrijanis\cite{MansSetEq} used to prove quantum query lower bound
for modified set equality problem. We will conclude that any
quantum query algorithm that solves the one-to-one set equality
problem either solves the collision problem or some other problem
that will be presented later. For the collision problem we have a
quantum query lower bound and for this other problem we will prove
it using Ambainis' adversary method. This implies lower bound for
the one-to-one set equality problem.

Unfortunately, since these reductions are probabilistic ones, but
Theorem~\ref{AThm} tells about ordinary functions, a lot of
technical work must be done to provide the correctness of the last
reduction. We will analyze properties of those reductions and
show, informally, that they are very similar (in sense of query
complexity).

There will be two kinds of reduction from the collision problem to
the set equality problem. Let $f$ denote r-to-one function which
the collision problem has in the input. From $f$ we will randomly
get two functions, $a$ and $b$. The both reductions will randomly
permutate range and domain of $f$ and divide domain into 2
disjoint halves. The first reduction will takes those halves of
domain as domains for functions $a$ and $b$. The second reduction
will take only the first half for both functions, just it will
make additional permutation over domain for both functions.

Informally, it is clear that both reductions makes "almost" equal
pair of functions $a$ and $b$ whenever $r$ is big "enough". We
will show that any quantum algorithm that can make distinction
between them must make "quite many" queries. On the other hand,
every algorithm for the set equality problem that don't make
distinction between them can be used to solve the collision
problem that is proved to be hard.

\section{Framework of the proof} \label{sec:Framework}
We have some $n$ and $1 < r < n$, such that $2|n$ and $r|n$. From
the conditions of the collision problem we have function $f:[n]
\rightarrow [N]$ with promise that $f$ is either one-to-one or
r-to-one. Let us choose random variables $\sigma \in SG[n],
\sigma_1 \in SG[n/2], \sigma_2 \in SG[n/2]$ and $\tau \in SG[N]$.

With \textbf{complementary} reduction we will denote the process
deriving functions $a$ and $b$ such that $a(i) =
\Gamma^{\sigma}_{\tau}(f)(i)$ and $b(i) =
\Gamma^{\sigma}_{\tau}(f)(n/2+i)$ for all $i \leq n/2$.

With \textbf{equivalent} reduction we will denote the process
deriving functions $a$ and $b$ such that $a(\sigma_1(i)) =
\Gamma^{\sigma}_{\tau}(f)(i)$ and $b(\sigma_2(i)) =
\Gamma^{\sigma}_{\tau}(f)(i)$ for all $i \leq n/2$.

\begin{Lemma} \label{lem:recog:colORfals}

For any quantum algorithm $\mathcal{A}$ that solves the set
equality problem with $T$ queries either there exists quantum
algorithm that solves r-to-one collision and makes $O(T)$ queries
or there exists quantum algorithm that makes distinction between
complimentary and equivalent reduction and makes $O(T)$ queries.
\end{Lemma}
\proof This tabular shows the acceptance probability of algorithm
$\mathcal{A}$ running on $a$ and $b$.

\begin{center}
\begin{tabular}{|c|c|c|}
  \hline
  reduction's type $\setminus$ function's $f$ type    & one-to-one & r-to-one \\ \hline
  complimentary                                       & $p^c_1>4/5$ & $p^c_2$ \\ \hline
  equivalent                                          & $p^e_1<1/5$ & $p^e_2$ \\\hline
\end{tabular}
\end{center}

There are two possibilities. If $p^e_2 \geq 2/5$ or $p^c_2 \leq
3/5$ then algorithm $\mathcal{A}$ can be used to solve the
collision problem. But if $p^e_2 < 2/5$ and $p^c_2 > 3/5$ then
algorithm $\mathcal{A}$ can be used to make distinction between
complimentary and equivalent reduction. \qed

In the next sections we will prove the following lemma:
\begin{Lemma} \label{lem:recog:falsif}

Any quantum algorithm $\mathcal{A}$ that makes distinction between
complimentary and equivalent reduction makes
$\Omega(\sqrt{\frac{r}{\log n}})$ queries.
\end{Lemma}

Choosing $r=n^{2/5}\log^{3/5} n$ Lemma~\ref{lem:recog:colORfals}
together with Lemma~\ref{lem:recog:falsif} and Theorem~\ref{shi}
will finish the proof of Theorem~\ref{thm:Main}.

\section{The lower bound of distinction, preparation}\label{sec:DistinctionPrep}
In this section we will start to prove
Lemma~\ref{lem:recog:falsif}. Informally, both reductions,
complementary and equivalent, make "quite similar" pairs of
functions. So we have to define what means "similar" and to proof
exactly how similar. Also, Theorem~\ref{AThm} deals with ordinary
input not distributions over inputs, therefore we will need to
formulate ordinary problem and reduce it to ours.

In this subsection we will investigate properties of both
reductions.

We will speak only about pairs of functions $(a,b)$ that can be
result of either complementary or equivalent reduction with
nonzero probability $p(a, b)>0$. We will investigate what pairs
$(a,b)$ can appear.

For any function $a$, which is in some pairs, let $INV(a):=(a_i |
0 \leq i \leq r):=(a_0,a_1,...,a_{r-1},a_r)$ denote the tuple
where $a_i$ is the number of image's elements $x \in [N]$, such
that cardinality of the set of preimages of $x$ is $i$, formally
$a_i := \#x(|a^{-1}(x)| = i)$ for $0 < i \leq r$ and
$a_0:=\frac{n}{r}-\sum\limits_{i=1}^r a_i$, where $\frac{n}{r}$ is
just the total count of images.

Let $INV(a, b)$ denote $(INV(a), INV(b))$. $INV(a, b)$ is quite
good way to describe the structure of some pair of functions $(a,
b)$ because of many reasons. Firstly, one can see, that $INV(a,
b)=INV(\Gamma_\tau^{\sigma_1}(a), \Gamma_\tau^{\sigma_2}(b))$ for
any pair of functions $(a, b)$ and any $\sigma_1 \in SG[n/2],
\sigma_2 \in SG[n/2]$ and $\tau \in SG[N]$.

Also, the probability for any pair $(a, b)$ to appear after
reduction $p(a, b)$ depends only on $INV(a, b)$. Moreover, if
there exists pairs of functions $(a_1, b_1)$ and $(a_2, b_2)$ such
that $INV(a_1,b_1)=INV(a_2,b_2)$ then there exists variables
$\sigma_1 \in SG[n/2], \sigma_2 \in SG[n/2]$ and $\tau \in SG[N]$
such that $(a, b)=(\Gamma_\tau^{\sigma_1}(a),
\Gamma_\tau^{\sigma_2}(b))$.

Now we will show, that, for any pair $(a, b)$, $INV(a)$ and
$INV(b)$ are closely related. For any $INV(a)=
(a_0,a_1,...,a_{r-1},a_r)$ let $\overline{INV}(a)$ denote the
tuple $(a_{r-i} | 0 \leq i \leq r):=(a_r,a_{r-1},...,a_1,a_0)$.

It is evident that for any pair of functions $(a,b)$ that occurs
after complementary reduction holds $INV(a) = \overline{INV}(b)$
but for any pair of functions $(a,b)$ that occurs after equivalent
reduction holds $INV(a) = INV(b)$.

We will use these facts to show that complementary and equivalent
reductions are quite similar, in other words, any functions $b_1$
and $b_2$ such that $INV(b_1) = \overline{INV}(b_2)$ differ in
many bits only with very small probability. So we will be able to
use Ambainis' Theorem~\ref{AThm} about bit's block flip to show
lower bound.

Let $a$ be any function that stand in some pair $(a,b)$ with
$INV(a) = (a_0, a_1, ..., a_r)$. Let $DISP(INV(a)) :=
\max\limits_{0 \leq i \leq r}(|i-r/2| : a_i>0)$.
\begin{Definition}\label{def:BAD}
    We say that $a$ is "bad" (and denote $BAD(a)$) if $DISP(INV(a)) > 15\sqrt{r \ln\frac{n}{r}}$.
\end{Definition}

Informally, $a$ is bad if there exists image such that after
reduction most of its preimages are in domain of either $a$ or
$b$.

It is easy to see that for any pair $(a, b)$ holds: $a$ is bad if
and only if $b$ is bad.

This Lemma shows, that the difference between complementary and
equivalent reductions is quite small:
\begin{Lemma}\label{lem:crRELATer}
    The sum $\sum\limits_{(a,b),BAD(a)} p(a, b)$
    is less than small constant if $r \gg \ln\frac{n}{r}$.
\end{Lemma}

\proof Let us see only the case when $(a,b)$ occurs after
complementary reduction. Let $f:n \rightarrow N$ be the function
before reduction. Let choose some fixed image $j$ of function $f$,
thus $j \in f([n])$. We say that $j$ is "bad" (denote by $BAD(j)$)
if $||a^{-1}(j)|-r/2|>15\sqrt{r \ln\frac{n}{r}}$. Let $p_j$ denote
the probability that $j$ is bad. It is easy to see that for all $j
\in f([n])$ $p_j$ is equal with some $p$. It is easy to see that
$BAD(a) \Leftrightarrow BAD(j)$ for some $j \in f([n])$. Therefore
probability for $a$ to be bad is less than $n/r*p$ where $n/r$ is
the total count of images. Now it remains only to show that $p \ll
r/n$.

There are two cases how $j$ can be bad, the first is that
$|a^{-1}(j)|$ is too big and the second is that $|a^{-1}(j)|$ is
too small. Obviously, that both of these cases holds with similar
probability. Let's count the probability that $|a^{-1}(j)|$ is too
big. Let enumerate all preimages of $j$ as $x_1, x_2, ..., x_r$.
Let $\chi'_i$ denote the random variable that is $1$ if $x_i$
become a member of domain of $a$ and $0$ elsewhere. Let
$\chi':=\chi'_1+\chi'_2+...+\chi'_r$. Thus we reach out for
$Pr[\chi'>E(\chi')+15\sqrt{r \ln\frac{n}{r}}]$, since
$E(\chi')=r/2$.

Let $\chi_i$ denote the random variable that is $1$ with
probability $1/2$ and $0$ with probability $1/2$. Let
$\chi:=\chi_1+...+\chi_r$. It is easy to see that for all $s \geq
r/2=E(\chi')=E(\chi)$ holds $Pr[\chi'>s] \leq Pr[\chi>s]$. Now we
can apply Chernoff's inequality $$Pr[\chi>(1+\epsilon)E(\chi)]
\leq e^{-\epsilon^2 E(\chi)/3}$$ if $0 \leq \epsilon \leq 1$. It
is easy to see that $E(\chi)=r/2$. Let $\epsilon :=
30\sqrt{\frac{\ln \frac{n}{r}}{r}}$. It is easy to see that
$\epsilon E(\chi)=15\sqrt{r \ln\frac{n}{r}}$. It remains to
evaluate the probability $e^{-\epsilon^2 E(\chi)/3}\ll r/n$ if $r
\gg \ln\frac{n}{r}$. \qed

\section{Proof's completing}
\label{sec:Compl}

In this section we will reduce the problem to distinguish between
complementary and equivalent reduction (with distribution over
input) to problem of the ordinary input.

\begin{Definition}\label{def:ComesFrom}
Let $n>0$ and $r \geq 2$ be integers such that $n|2$ and $n|r$.
Let $a:n/2\rightarrow N$ and $b:n/2\rightarrow N$ be functions
given by an oracle, such that the pair $(a,b)$ can occur after
complementary or equivalent reduction. $INV(a)$ is known and it is
promised that $a$ is not bad. Let $ComesFrom$ problem denote the
problem to decide whether the pair $(a,b)$ occurred after
complementary or equivalent reduction.
\end{Definition}

\subsection{Reduction}
\label{sec:Compl:Reduction}

\label{lem:distrVSord}
\begin{Lemma}
    If there exists quantum algorithm $\mathcal{A}$ that
    makes distinction between complimentary and equivalent reduction
    with $T$ queries then there exists quantum algorithm $\mathcal{A}'$
    that solves ComesFrom problem with $O(T)$ queries.
\end{Lemma}
\proof

Firstly, we can ignore all pairs $(a, b)$ that have $BAD(a)$
because they appear with very small probability (
Lemma~\ref{lem:crRELATer}). If we want to improve the probability
we can just repeat $\mathcal{A}$ several times.

Secondly, without loss of generality, we can assume that the
accepting probability of $\mathcal{A}$ depends only on $INV(a,
b)$. If not, we can modify  algorithm $\mathcal{A}$, such that it
choose random variables $\sigma_1 \in SG[n/2], \sigma_2 \in
SG[n/2]$ and $\tau \in SG[N]$ at the beginning and further just
deal with pair of functions $(\Gamma_\tau^{\sigma_1}(a),
\Gamma_\tau^{\sigma_2}(b))$.

Thirdly, since $\mathcal{A}$ makes distinction between
complimentary and equivalent reduction and for any pair of
functions $(a,b)$ depends only on $INV(a, b)$, there exists some
$I := INV(a)$ such that $\mathcal{A}$ makes distinction between
$(a, b_1)$ such that $INV(b_1) = INV(a) = I$ and $(a, b_2)$ such
that $\overline{INV}(b_2) = INV(a) = I$ for any function $b_1$ and
$b_2$.

It follows, that for this particular $I$ we can solve ComesFrom
problem using algorithm $\mathcal{A}$.

 \qed

\subsection{Lower bound for the $ComesFrom$ problem}
\label{sec:Compl:ComesFrom}
\begin{Lemma}\label{lem:ComesFrom}
    Any quantum algorithm $\mathcal{A}$ that solves the $ComesFrom$
    problem makes $\Omega\left(\sqrt{\frac{r}{\log n}}\right)$ queries.
\end{Lemma}
\proof Let $I = (a_0, ..., a_r)$ denote the known $INV(a)$. We
will use Theorem~\ref{AThm} to prove lower bound quite similarly
to Ambainis'~\cite{Ambainis} proof about lower bound for counting.
Let $X$ be the set of all $(a, b)$ such that $INV(b) = INV(a) = I$
and let $Y$ be the set of all $(a, b)$ such that
$\overline{INV}(b) = INV(a) = I$.

Let $\Psi := \Psi(I) := \sum\limits_{i: a_i > r/2}a_i-r/2$. Since
$a$ is not bad, it implies that $\Psi = O(\frac{n}{r} \sqrt{r \log
n})$. $2\Psi$ is just the number of points that must be changed to
to switch from $INV(b)$ to $\overline{INV}(b)$. Let $\Phi :=
\frac{(2\Psi)!}{\prod\limits_{i: a_i < r/2}(r-2a_i)!}$.

Let $R$ be the set of all $((a, b_1), (a, b_2))$ such that $b_1$
differs from $b_2$ exactly in $2\Psi$ points and $INV(b_1) =
\overline{INV}(b_2)$. Then, $m = m' = C^{2\Psi}_{n/4+\Psi}\Phi$
and $l = l' = C^{2\Psi-1}_{n/4+\Psi-1} \Phi$. Therefore,
$$\frac{mm'}{ll'} = \left(\frac{C^{2\Psi}_{n/4+\Psi}\Phi}{C^{2\Psi-1}_{n/4+\Psi-1}\Phi}\right)^2=
\left(\frac{(n/4+\Psi)!(2\Psi-1)!(n/4-\Psi)!}{(2\Psi)!(n/4-\Psi)!(n/4+\Psi-1)!}\right)^2=
\left(\frac{n/4+\Psi}{2\Psi}\right)^2=$$
$$=\left(\frac{n}{8\Psi}+\frac{1}{2}\right)^2 =
\Omega\left(\left(\frac{n}{\Psi}\right)^2\right)=
\Omega\left(\frac{r}{\log n}\right)$$

Now we can apply Theorem~\ref{AThm} and get that any quantum
algorithm makes $\Omega\left(\sqrt{\frac{r}{\log n}}\right)$
queries. \qed

\section{Conclusion}\label{conclusion}
We showed a polynomial quantum query lower bound for the set
equality problem. It was done by reduction. Arguments that allowed
reduction was very specific to the set equality problem. It would
be nice to find some more general approach to find quantum query
lower bounds for this and other similar problems. Also, it would
be fine to make smaller difference between quantum lower and upper
bounds for the set equality problem.

\section{Acknowledgments}
I am very thankful to Andris Ambainis about introduction and
discussions about this problem, quantum lower bounds and quantum
computation, and useful comments about this paper, to Inga \u
Cerniauskait\.e about checking my spelling, to Andrejs Dubrovskis
for discussions after my first paper about quantum query lower
bound for the set equality problem, to Yufan Zhu about pointing
out to mistake in previous version of Lemma's~\ref{lem:ComesFrom}
proof and to anonymous referees whose comments helped me to
improve presentation of the result.

\end{document}